\documentclass[twoside,fleqn]{article}
\usepackage{espcrc2}
\usepackage{graphicx}
\newcommand{\be}{\begin{equation}}
\newcommand{\ee}{\end{equation}}
\newcommand{\chiPT}{$\chi$PT}

\title{ 
\vspace{-2.6cm}
\hfill \rm \null \hfill
\hbox{\normalsize ADP-03-128/T564} \\
\vspace{-2mm}
\hfill \hbox{\normalsize DESY 03-210} \\
\vspace{1.65cm}
Electromagnetic Form Factors with FLIC fermions}

\author{J.~M.~Zanotti\address[CSSM]{Special Research Center for the
    Subatomic Structure of Matter, and          \\
    Department of Physics, University of Adelaide Adelaide SA 5005
    Australia}\address[DESY]{John von Neumann-Institut f\"ur Computing
    NIC, \\
    Deutsches Elektronen-Synchrotron DESY, D-15738 Zeuthen, Germany}
  \thanks{Presented by J.~M.~Zanotti at LHP '03, Cairns, Australia},
  S.~Boinepalli\addressmark[CSSM], D.~B.~Leinweber\addressmark[CSSM],
  A.~G.~Williams\addressmark[CSSM] and J.~B.~Zhang\addressmark[CSSM] }

\begin{document}

\begin{abstract}
  The Fat-Link Irrelevant Clover (FLIC) fermion action provides a new
  form of nonperturbative ${\cal O}(a)$ improvement and allows
  efficient access to the light quark-mass regime.  FLIC fermions
  enable the construction of the nonperturbatively ${\cal
  O}(a)$-improved conserved vector current without the difficulties
  associated with the fine tuning of the improvement coefficients.
  The simulations are performed with an ${\cal O}(a^2)$ mean-field
  improved plaquette-plus-rectangle gluon action on a $20^3\times 40$
  lattice with a lattice spacing of 0.128 fm, enabling the first
  simulation of baryon form factors at light quark masses on a large
  volume lattice.
  Magnetic moments, electric charge radii and magnetic radii are
  extracted from these form factors, and show interesting chiral
  nonanalytic behavior in the light quark mass regime.
\end{abstract}

\maketitle

%\vspace{3mm}
%PACS number(s): 11.15.Ha, 12.38.Gc, 12.38.Aw

%\newpage

%%%%%%%%%%%%%%%%%%%%%%%%%%%%%%%%%%%%%%%%%%%%%%%%%%%%%%%%%%%%%%%%%%%%%%%%%%%
\section{INTRODUCTION}

The magnetic moments of baryons have been identified as providing an
excellent opportunity for the direct observation of chiral nonanalytic
behavior in lattice QCD, even in the quenched approximation
\cite{Leinweber:2001jc,Savage:2001dy,Leinweber:2002qb}.  This paper
will present results for baryon electromagnetic structure in which the
chiral nonanalytic behaviour predicted by quenched chiral perturbation
theory is observed in the numerical simulation results.

The numerical simulations of the electromagnetic form factors
presented here are carried out using the Fat Link Irrelevant Clover
(FLIC) fermion action \cite{FATJAMES,Leinweber:2002bw} in which the
irrelevant operators introduced to remove fermion doublers and lattice
spacing artifacts \cite{Bilson-Thompson:2002jk} are constructed with
smoothed links.  These links are created via APE smearing
\cite{ape}. On the other hand, the relevant operators surviving in the
continuum limit are constructed with the original untouched links
generated via standard Monte Carlo techniques.

FLIC fermions provide a new form of nonperturbative ${\cal O}(a)$
improvement \cite{Leinweber:2002bw,inPrep} where near-continuum
results are obtained at finite lattice spacing.  Access to the light
quark mass regime is enabled by the improved chiral properties of the
lattice fermion action \cite{inPrep}.
%The magnitude of additive mass
%renormalizations is suppressed \cite{inPrep} which otherwise can
%lead to singular behavior in the propagators as the quarks become
%light.
%
The ${\cal O}(a)$-improved conserved vector current
\cite{Martinelli:ny} is used.  Nonperturbative improvement is achieved
via the FLIC procedure where the terms of the Noether current having
their origin in the irrelevant operators of the fermion action are
constructed with mean-field improved APE smeared links.
The use of links in which short-distance fluctuations have been
removed simplifies the determination of the coefficients of the
improvement terms in both the action and its associated conserved
vector current.  Perturbative renormalizations are small for smeared
links and are accurately
accounted for by small mean-field improvement corrections.  Hence, we
are able to determine the form factors of octet and decuplet baryons
with unprecedented accuracy.

\vspace*{-0.1cm}
%%%%%%%%%%%%%%%%%%%%%%%%%%%%%%%%%%%%%%%%%%%%%%%%%%%%%%%%%%%%%%%%%%%%%%%%%%%
\section{LATTICE TECHNIQUES}

\subsection{Gauge and Quark Actions}
\label{FLinks}

The simulations are performed using a mean-field ${\cal
O}(a^2)$-improved Luscher-Weisz \cite{Luscher:1984xn} gauge action on
a $20^3 \times 40$ lattice with a lattice spacing of 0.128 fm as
determined by the Sommer scale $r_0=0.50$ fm.  We use a minimum of 255
configurations and the error analysis is performed by a third-order,
single-elimination jackknife.

For the quark fields, we use the Fat-Link Irrelevant Clover fermion
action \cite{FATJAMES}. Fat links 
%\cite{DeGrand:1998jq,DeGrand:1999gp} 
are created using APE smearing \cite{ape}.
The smearing procedure replaces a link, $U_{\mu}(x)$, with a
sum of the link and $\alpha$ times its staples
\begin{eqnarray}
U_{\mu}(x)\ &\rightarrow&\ U_\mu^{\rm FL}(x)\ =\
(1-\alpha) U_{\mu}(x) \\
&&\hspace*{-12mm} + \frac{\alpha}{6}\sum_{\nu=1 \atop \nu\neq\mu}^{4}
  \Big[ U_{\nu}(x)
        U_{\mu}(x+\nu a)
        U_{\nu}^{\dag}(x+\mu a)                         \nonumber \\
\mbox{} 
&&\hspace*{-12mm} + U_{\nu}^{\dag}(x-\nu a)
        U_{\mu}(x-\nu a)
        U_{\nu}(x-\nu a +\mu a)
  \Big] \, , \nonumber
\end{eqnarray} 
followed by projection back to SU(3). We select the unitary matrix
$U_{\mu}^{\rm FL}$ which maximizes
$$
{\cal R}e \, {\rm{tr}}(U_{\mu}^{\rm FL}\, U_{\mu}'^{\dagger})\,
$$
by iterating over the three diagonal SU(2) subgroups of SU(3).
We repeat this procedure of smearing followed immediately by projection $n$ times.
We select a smearing fraction of
$\alpha = 0.7$ (keeping 0.3 of the original link) and iterate the
process six times \cite{Bonnet:2000dc}. 
The mean-field improved FLIC \cite{FATJAMES} action now becomes
\be
S_{\rm SW}^{\rm FL}
= S_{\rm W}^{\rm FL} - \frac{iC_{\rm SW} \kappa r}{2(u_{0}^{\rm FL})^4}\
             \bar{\psi}(x)\sigma_{\mu\nu}F_{\mu\nu}\psi(x)\ ,
\ee
where $F_{\mu\nu}$ is constructed using fat links, and where the
mean-field improved Fat-Link Irrelevant Wilson action is
\begin{eqnarray}
S_{\rm W}^{\rm FL}
 &=&  \sum_x \bar{\psi}(x)\psi(x) \\
&+& \kappa \sum_{x,\mu} \bar{\psi}(x)
    \bigg[ \gamma_{\mu}
      \bigg( \frac{U_{\mu}(x)}{u_0} \psi(x+\hat{\mu}) \nonumber\\
& & \qquad - \frac{U^{\dagger}_{\mu}(x-\hat{\mu})}{u_0} \psi(x-\hat{\mu})
      \bigg)                                            \nonumber\\
&-& r \bigg(
          \frac{U_{\mu}^{\rm FL}(x)}{u_0^{\rm  FL}} \psi(x+\hat{\mu}) \nonumber\\
& & \qquad + \frac{U^{{\rm FL}\dagger}_{\mu}(x-\hat{\mu})}{u_0^{\rm FL}}
          \psi(x-\hat{\mu})
      \bigg)
    \bigg]\ , \nonumber
\end{eqnarray}
with $\kappa = 1/(2m+8r)$. We take the standard value $r=1$.
The $\gamma$-matrices are hermitian and
$\sigma_{\mu\nu} = [\gamma_{\mu},\ \gamma_{\nu}]/(2i)$.

For fat links, the mean link $u_0 \approx 1$, enabling the use of
highly improved definitions of the lattice field strength tensor,
$F_{\mu\nu}$ \cite{Bilson-Thompson:2002jk}.  In particular, we employ
an ${\cal O}(a^4)$-improved definition of $F_{\mu\nu}$ in which the
standard clover-sum of four $1 \times 1$ Wilson loops lying in the
$\mu ,\nu$ plane is combined with $2 \times 2$ and $3 \times 3$ Wilson
loop clovers.  Moreover, mean-field improvement of the coefficients of
the clover and Wilson terms of the fermion action is sufficient to
accurately match these terms and eliminate ${\cal O}(a)$ errors from
the fermion action \cite{Leinweber:2002bw,inPrep}. 

% As the conserved vector current has its origin in
% the fermion action, mean-field improvement of the irrelevant operator
% terms is also sufficient to accurately remove ${\cal O}(a)$ errors,
% providing a nonperturbatively ${\cal O}(a)$-improved conserved vector
% current.

Previous work \cite{FATJAMES,WASEEM} has shown that the FLIC fermion
action has extremely impressive convergence rates for matrix
inversion, which provides great promise for performing cost effective
simulations at quark masses closer to the physical values.  Problems
with exceptional configurations have prevented such simulations in the
past.
The ease with which one can invert the fermion matrix using FLIC
fermions leads us to attempt simulations down to light quark masses
corresponding to $m_{\pi} / m_{\rho} = $ 0.35 in an attempt to
reveal chiral non-analytic behaviour in baryon magnetic moments.

A fixed boundary condition at $t=0$ is used for the fermions and
gauge-invariant Gaussian smearing \cite{Gusken:qx,Zanotti:2003fx} in
the spatial dimensions is applied at the source at $t=8$ to increase
the overlap of the interpolating operators with the ground state while
suppressing excited state contributions.  

\subsection{Improved Conserved Vector Current}

For the construction of the ${\cal O}(a)$-improved conserved vector
current, we follow the technique proposed by Martinelli {\it et al.}
\cite{Martinelli:ny}.  The standard conserved vector current for
Wilson-type fermions is derived via the Noether procedure
\begin{eqnarray}
j_\mu^{\rm C} &\equiv& \frac{1}{4}\bigl[\overline{\psi}(x) (\gamma_\mu -
r)U_\mu(x) \psi(x+\hat{\mu}) \nonumber \\ 
&+& \overline{\psi}(x+\hat{\mu}) (\gamma_\mu + r)U_\mu^\dagger(x)
\psi(x) \nonumber \\
&+& (x\rightarrow x-\hat{\mu})\bigr] .
\label{conserved}
\end{eqnarray}
The ${\cal O}(a)$ improvement term is also derived from the fermion
action and is constructed in the form of a total four-divergence,
preserving charge conservation.  The ${\cal O}(a)$-improved conserved
vector current is
\be
j_\mu^{\rm CI} \equiv j_\mu^{\rm C} (x) + \frac{r}{2} C_{CVC}\, a \sum_\rho
\partial_\rho \bigl( \overline{\psi}(x) \sigma_{\rho\mu}\psi(x)\bigr)
\, ,
\label{impconserved}
\ee
where $C_{CVC}$ is the improvement coefficient for the conserved
vector current and we define
\be
\partial_\rho \bigl( \overline{\psi}(x) \psi(x)\bigr) \equiv
\overline{\psi}(x) \bigl( \overleftarrow{\nabla}_\rho +
\overrightarrow{\nabla}_\rho \bigr) \psi(x)\, ,
\ee
where the forward and backward derivatives are defined as
\begin{eqnarray*}
\overrightarrow{\nabla}_{\mu}\psi(x) &=&
\frac{1}{2a} \left [
U_{\mu}(x)\, \psi(x+\hat{\mu}) \right . \\  
&&\qquad\qquad
\left . - U^{\dagger}_{\mu}(x-\hat{\mu}) \, \psi(x-\hat{\mu})
\right ] \, ,\\
\overline{\psi}(x)\overleftarrow{\nabla}_{\mu} &=& \frac{1}{2a} \left [
\overline\psi(x+\hat{\mu})\,  U_{\mu}^\dagger(x) \right . \\
&&\qquad\qquad
\left . - \overline\psi(x-\hat{\mu}) \, U_{\mu}(x-\hat{\mu})
\right ]\, .
\end{eqnarray*}

The terms proportional to the Wilson parameter $r$ in
Eq.~(\ref{conserved}) and the four-divergence in
Eq.~(\ref{impconserved}) have their origin in the irrelevant operators
of the fermion action and vanish in the continuum limit.
Nonperturbative improvement is achieved by constructing these terms
with fat-links.  As we have stated, perturbative corrections are small
for fat-links and the use of the tree-level value for $C_{CVC} = 1$
together with small mean-field improvement corrections ensures that
${\cal O}(a)$ artifacts are accurately removed from the vector
current.  This is only possible when the current is constructed with
fat-links.  Otherwise, $C_{CVC}$ needs to be appropriately tuned to
ensure all ${\cal O}(a)$ artifacts are removed.

\subsection{Lattice Three-Point Functions}

The technique used for
constructing the three-point functions follows the procedure outlined
in detail in Refs.~\cite{Leinweber:1990dv,Leinweber:1992hy}. In
particular, we use the sequential source technique at the current
insertion.  Correlation functions are made purely real with exact
parity through the consideration of $U$ and $U^*$ link configurations.
Electric and magnetic form factors are extracted by constructing
ratios of two- and three-point functions
for a baryon, $B$
\begin{eqnarray}
\lefteqn{
  R(t_2,t_1; \vec{p'}, \vec{p}; { \Gamma}; { \mu} ) = 
} \nonumber \\
%   R(t_2,t_1; \vec{p'}, \vec{p}; \Gamma, \Gamma^\prime; \mu ) = } \\
  & \left ( {\big <  G^{B j^\mu B} (t_2,t_1; \vec{p'}, \vec{p};
\Gamma) \big >
\big <  G^{ B j^\mu B} (t_2,t_1; -\vec{p}, -\vec{p'};
\Gamma) \big >
\over
\big < G^{BB} (t_2; \vec{p'}; \Gamma_4 ) \big >
\big < G^{BB} (t_2; -\vec{p}; \Gamma_4 ) \big > } 
\right )^{1/2} \nonumber \\
%\begin{equation}
&\simeq 
  \left ( { E_p + M \over 2 E_p } \right )^{1/2}
  \left ( { E_{p'} + M \over 2 E_{p'} } \right )^{1/2}
 { \overline R}(\vec{p'}, \vec{p}; \Gamma; \mu ) 
\label{3to2ratio}
%   \overline R(\vec{p'}, \vec{p}; \Gamma, \Gamma^\prime; \mu )  
%\end{equation}
\end{eqnarray}
where $0\gg t_1 \gg t_2$ are the current insertion and baryon sink
time slices respectively; $\vec{p}$ and $\vec{p'}$ are the baryon
momentum before and after current insertion respectively; and $\Gamma$
is a $4\times4$ dirac matrix with $\Gamma_4 =
\frac{1}{4}(1+\gamma_4)$. The Sachs form of the electromagnetic form
factors,
\begin{eqnarray*}
{\cal G}_E(q^2) &=& F_1(q^2) - \frac{q^2}{(2M)^2} F_2(q^2)\, ,\\
{\cal G}_M(q^2) &=& F_1(q^2) + F_2(q^2)\, ,
\end{eqnarray*}
are obtained from $\overline R$ by
\begin{eqnarray*}
 {\overline R}(\vec{q},\vec{0}; \Gamma_4, 4) &=& {\cal
%  \overline R(\vec{q},\vec{0}; \Gamma_4, \Gamma_4, 4) &=& {\cal
  G}_E(q^2)\ , \\
 {\overline R}(\vec{q},\vec{0}; \Gamma_j, k) &=&
%  \overline R(\vec{q},\vec{0}; \Gamma_j, \Gamma_4, k) &=&
       \frac{ {\cal G}_M(q^2) \; | \epsilon_{ijk} q^i | }
            { (E_q + M) } \ , \\
 {\overline R}(\vec{q},\vec{0}; \Gamma_4, k) &=&
       \frac{ {\cal G}_E(q^2) \; | q^k | }{ (E_q + M) } \ .
\end{eqnarray*}
We simulate at the smallest finite $q^2$ available on our lattice,
$ {\vec q = \frac{2 \pi}{a L} \widehat x}$.
We insert our improved conserved vector current at $t_1=14$. Since
${\overline R}$ is independent of the baryon sink time slice, we
calculate the ratio in Eq.~(\ref{3to2ratio}) for a range of sink
times. Using a covariance-matrix-based $\chi^2/N_{dof}$, we
independently select suitable fitting windows for both electric and
magnetic form factors.

\subsection{Correlation Function Analysis}

In selecting the most appropriate Euclidean time window for fitting, we
consider the correlated $\chi^{2}/{\rm dof}$ as given by
\begin{eqnarray}
\frac{\chi^{2}}{\rm dof} &=& \frac{1}{N_{t}-M} \sum_{i=1}^{N_{t}} 
\sum_{j=1}^{N_{t}} \nonumber \\
&& (y(t_{i})-k) \, C^{-1}(t_{i},t_{j}) \, (y(t_{j})-k) \, , 
\end{eqnarray}
where, $M$ ($=1$) is the number of parameters to be fitted,
$N_{t}$ is the number of time slices considered, $y(t_{i})$ is the
configuration average value of the dependent variable at time $t_{i}$
that is being fitted to a constant value $k$, and $C(t_{i},t_{j})$ is
the covariance matrix. The elements of the covariance matrix are
estimated via the jackknife method,
\begin{eqnarray*}
C(t_{i},t_{j}) &=& \frac{N_{c}-1}{N_{c}} \sum_{m=1}^{N_{c}} \nonumber \\
&&\left ( \overline{y_{m}}(t_{i}) - \overline{\overline{y}}(t_{i})
\right ) \,  \left ( \overline{y_{m}}(t_{j}) -
\overline{\overline{y}}(t_{j}) \right ) \, ,
\end{eqnarray*} 
where, $N_{c}$ is the total number of configurations,
$\overline{y_{m}}(t_{i})$ is the jackknife ensemble average of the
system after removing the $\mathit{m_{th}}$
configuration. $\overline{\overline{y}}(t_{i})$ is the average of all
such jackknife averages, given by
\begin{equation}
\overline{\overline{y}}(t_{i})=\frac{1}{N_{c}}\sum_{m=1}^{N_{c}}
\overline{y_{m}}(t_{i}).
\end{equation}

\begin{figure}[t]
\begin{center}
\hspace*{-0.5cm}
{\includegraphics[height=\hsize,angle=90]{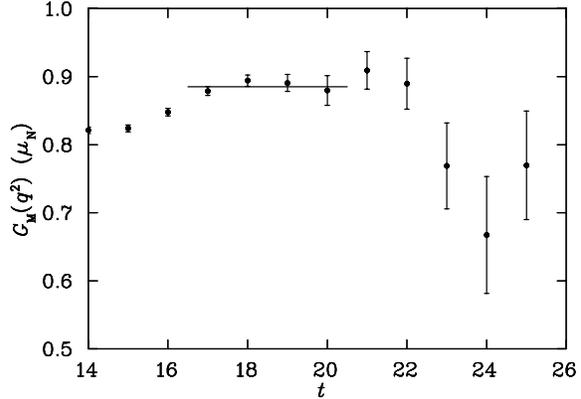}}
\vspace*{-0.5cm}
\caption{Contribution from a $u$ quark of unit charge to the proton
magnetic form factor.  The quark mass for this correlation functions
corresponds to $m_\pi^2 \simeq 0.3\ {\rm GeV}^2$. }
\label{mformfactor}
\end{center}
\end{figure}

Table \ref{table:1} shows the contribution from a $u$ quark of unit
charge to the proton magnetic form factor in nuclear magnetons
($\mu_N$).  The uncertainty and the $\chi^{2}/dof$ are quoted for
various time-fitting windows.  The principal value of the fitted
parameter remains fairly constant over different time windows, but the
$\chi^{2}/dof$ shows a marked decrease as we move from the time window
16-18 to 17-20.  Figure \ref{mformfactor} illustrates these lattice
simulation results obtained from Eq.~(\ref{3to2ratio}) for one of the
lighter quark masses.  The horizontal line indicates the best fit in
the time window 17-20 which provides the most acceptable $\chi^{2}/dof
\sim 1$.

\begin{table}[t]
\caption{Fit results for the contribution of a $u$ quark of unit
charge to the proton magnetic form factor ($\mu_N$) obtained for
various time-fitting windows.}
\label{table:1}
\begin{center}
\begin{tabular}{lccc}                                  
\hline
\noalign{\smallskip}
Time    & Fit Value   & Uncertainty   & $\chi^{2}/{\rm dof}$    \\  \hline
\noalign{\smallskip}
16-18   & 0.865       & 0.022          & 7.891                   \\
16-19   & 0.867       & 0.024          & 6.304                   \\
16-20   & 0.868       & 0.025          & 4.731                   \\
17-19   & 0.884       & 0.03           & 1.281                   \\
17-20   & 0.885       & 0.033          & 0.855                   \\
18-20   & 0.893       & 0.042          & 0.047                   \\  \hline
\end{tabular}
\end{center}
\end{table}

\section{RESULTS}
\label{discussion}

\begin{figure}[t]
\begin{center}
\hspace*{-0.5cm}
{\includegraphics[height=1.25\hsize,angle=270]{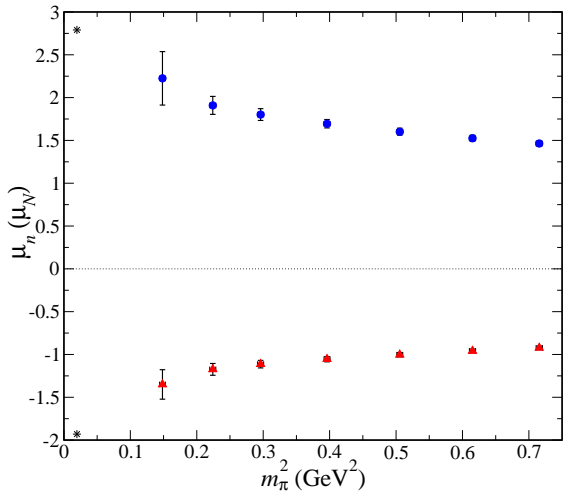}}
\vspace*{-0.5cm}
\caption{FLIC fermion simulation results for the magnetic moments of
  the proton (upper) and neutron (lower) in
  quenched QCD. The experimental values are given by asterisks}
\label{PNmoments}
\end{center}
%\vspace{-1.0cm}
\end{figure}

Figure \ref{PNmoments} displays FLIC fermion simulation results for
the magnetic moments of the proton and neutron in quenched QCD.  The
moments are obtained using the empirical fact that the $Q^2$
dependence of the electric and magnetic form factors of the proton
over the range $0 \to 0.23\ {\rm GeV}^2$ are nearly equivalent.
At heavy quark masses we note that the magnetic moments of both
nucleons display approximately linear behaviour when plotted as a
function of $m_\pi^2$.  Simulations at moderately heavier quark masses
are expected to reveal a Dirac moment behavior $\propto~1/m_q \sim
1/m_\pi^2$.  As one approaches the light quark mass regime,
we find evidence of non-analytic behaviour in the nucleon magnetic
moments as predicted by quenched \chiPT\
\cite{Leinweber:2001jc,Leinweber:2002qb}.
In fact, if we were to flip the sign of the neutron magnetic moment,
we would find that the proton and neutron have a very similar
behaviour as a function of $m_\pi^2$ in the light quark mass regime as
predicted by the leading non-analytic contributions of quenched \chiPT.

\begin{figure}
\begin{center}
\hspace*{-0.5cm}
{\includegraphics[height=1.25\hsize,angle=270]{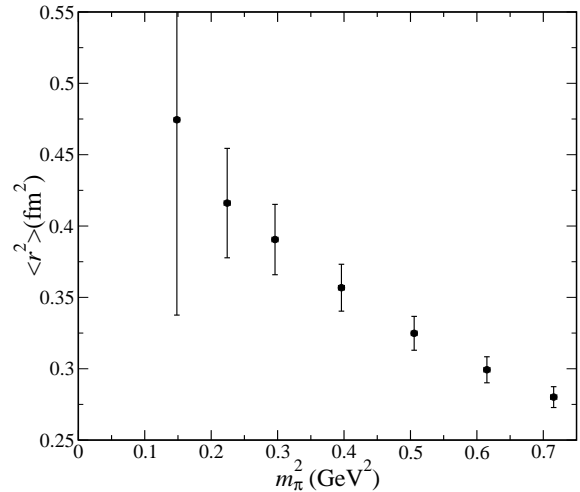}}
\vspace*{-0.5cm}
\caption{FLIC fermion simulation results for the charge radius of the
  proton in quenched QCD. }
\label{chargeRadii}
\end{center}
%\vspace{-1.0cm}
\end{figure}

\begin{figure}
\begin{center}
\hspace*{-0.5cm}
{\includegraphics[height=1.25\hsize,angle=270]{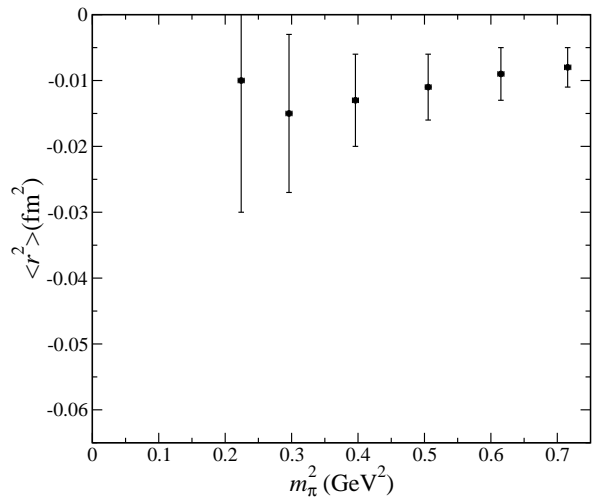}}
\vspace*{-0.5cm}
\caption{FLIC fermion simulation results for the charge radius of the
  neutron in quenched QCD. }
\label{NchargeRadii}
\end{center}
%\vspace{-1.0cm}
\end{figure}

Figures \ref{chargeRadii} and \ref{NchargeRadii} display results for
the proton and neutron charge radii, respectively, obtained from a
dipole form factor ansatz.  For the proton some curvature
is emerging as the chiral limit is approached.
The FLIC fermion simulations reveal a small, negative value for the
calculation of the quenched neutron charge radius at large and intermediate
quark masses before the signal is lost at the lightest two quark masses. This
result confirms the earlier result \cite{Leinweber:1990dv} that the
two $d$ quarks have a larger charge radius than the $u$ quark within
the neutron.  This is also in agreement with quark model calculations
\cite{Carlitz:bd}.

\begin{figure}
\begin{center}
\hspace*{-0.5cm}
{\includegraphics[height=1.25\hsize,angle=270]{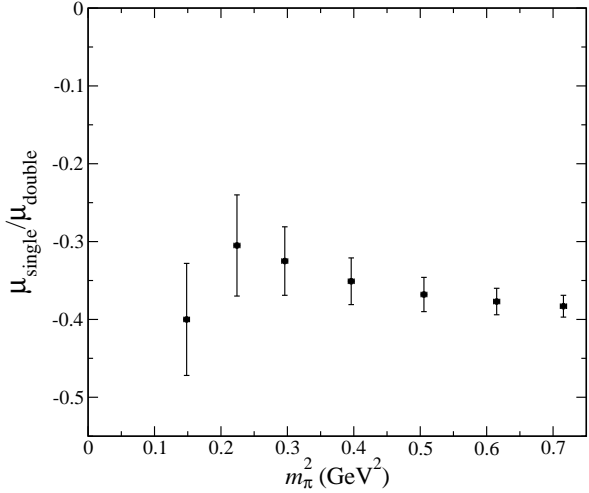}}
\vspace*{-0.5cm}
\caption{Ratio of singularly to doubly represented quark magnetic
  contributions in the nucleon for quarks of unit charge. }
\label{udMagRatio}
\end{center}
%\vspace{-1.0cm}
\end{figure}

We also perform a calculation of the individual quark sector
contributions to the magnetic moments.  
%In the simple quark model in
%the SU(2)-flavour limit, the ratio of the quark sector contributions
%in the proton is $\mu_u/\mu_d = -1/2$. 
In Fig.~\ref{udMagRatio} we show results for a calculation in quenched
QCD of the ratio of singly to doubly represented quark magnetic
contributions in the nucleon for quarks of unit charge. We see that
our results deviate significantly from the simple quark model
prediction of $-1/2$.

\begin{figure}
\begin{center}
\hspace*{-0.5cm}
{\includegraphics[height=1.25\hsize,angle=270]{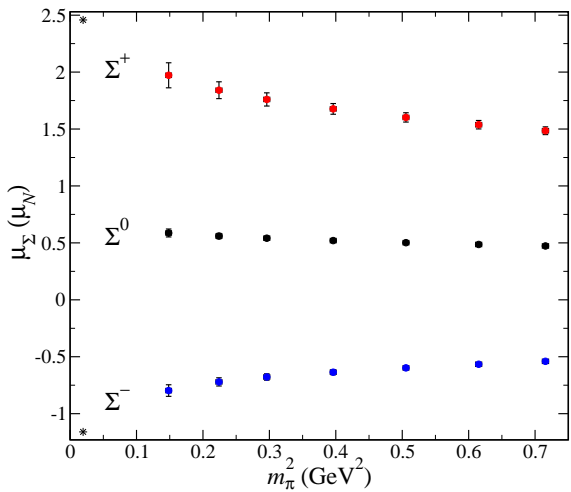}}
\vspace*{-0.5cm}
\caption{FLIC fermion simulation results for the magnetic moment of the
  octet $\Sigma$ baryons in quenched QCD. }
\label{sigmaMM}
\end{center}
%\vspace{-1.0cm}
\end{figure}

Figure \ref{sigmaMM} shows the results for a calculation of the octet
$\Sigma$ baryons. We note the level ordering of the three $\Sigma$
baryons, with the positive and negative charge states both
approaching the experimental values at light quark masses. We also see
that our quenched lattice calculation of the magnetic moment of the
neutral $\Sigma$ baryon predicts a value in the range $0.5-0.7\mu_N$,
although we first need to take into account the appropriate chiral
extrapolation and quenching effects
\cite{DerekLHP03} before an accurate
prediction can be made. Our results reveal a significant amount of
nonanalytic behaviour for the $\Sigma^+$, which is predicted by
quenched \chiPT\ \cite{Leinweber:2002qb}. Q\chiPT\ also predicts a small
amount of curvature for the $\Sigma^0$ and $\Sigma^-$ and we confirm
this prediction \cite{Leinweber:2002qb}. 

\begin{figure}
\begin{center}
\hspace*{-0.5cm}
{\includegraphics[height=1.25\hsize,angle=270]{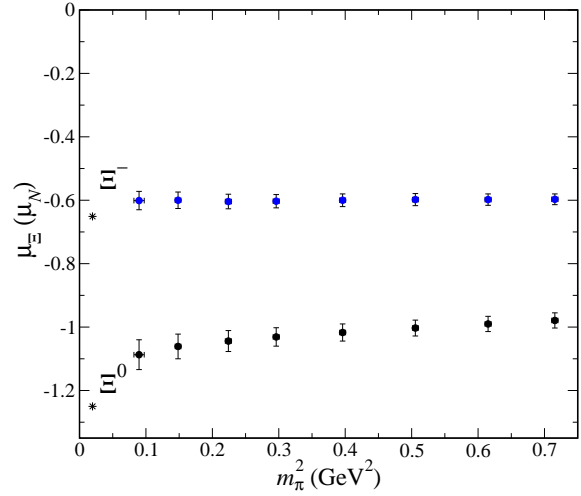}}
\vspace*{-0.5cm}
\caption{FLIC fermion simulation results for the magnetic moment of the
  octet $\Xi$ baryons in quenched QCD. }
\label{xiMM}
\end{center}
%\vspace{-1.0cm}
\end{figure}

We have also performed a calculation of the magnetic moment of the
octet $\Xi$ baryons in quenched QCD and these results are shown in
Fig.~\ref{xiMM}. Here we see an improved signal for the magnetic
moment of the $\Xi$ compared to the other baryons at light quark
masses due to the presence of two strange quarks. We note the splitting
between the neutral and negative charged baryons is close to the
experimentally observed value, even in the quenched approximation.
Predictions from quenched chiral perturbation theory
\cite{Leinweber:2002qb} suggest that there should be no nonanalytic
behaviour for the $\Xi^-$ and a small amount of downward curvature for
the $\Xi^0$. Our results confirm these predictions.

% \begin{table}
% \begin{center}
% \caption{$|\mu(7)-\mu(5)/\mu(5)-\mu(1)|$ for the octet baryons, where
%   $\mu(n)$ is the magnetic moment calculated at the $n$th quark mass
%   with $n=1$ being the heaviest quark mass.
%       \label{slopes}}
% \vspace*{0.5cm}
% \begin{tabular}{ccccccc}
% \hline \vspace*{0.9mm}
% $p$ & $n$ & $\Sigma^+$ & $\Sigma^0$ & $\Sigma^-$ &
% $\Xi^0$ & $\Xi^-$ \\ \hline
%  0.5 & 0.48 & 0.42 & 0.39 & 0.45 & 0.34 & 0 \\
% \hline
% \end{tabular}
% %\vspace*{0.5cm}
% \end{center}
% \end{table}

\begin{figure}
\begin{center}
\hspace*{-0.5cm}
{\includegraphics[height=1.25\hsize,angle=270]{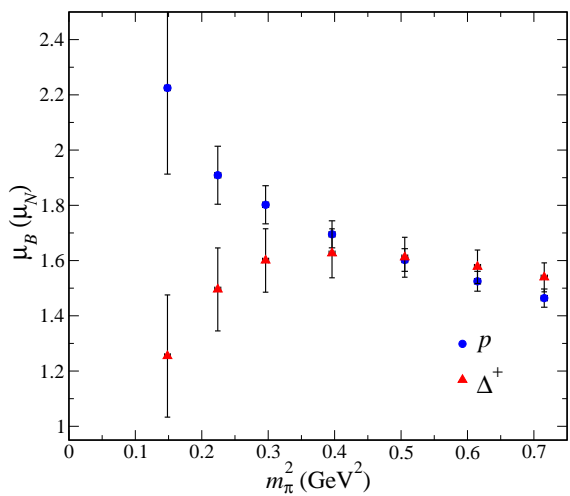}}
\vspace*{-0.5cm}
\caption{FLIC fermion simulation results for the magnetic moments of
  the proton ($\circ$) and $\Delta^+$ resonance ($\triangle$) in
  quenched QCD. }
\label{PDmoments}
\end{center}
%\vspace{-1.0cm}
\end{figure}

Figure \ref{PDmoments} displays FLIC fermion simulation results for the
magnetic moments of the proton and $\Delta^+$ resonance in quenched QCD.
At large pion masses, the $\Delta$ moment is enhanced relative to the
proton moment in accord with earlier lattice QCD results
\cite{Leinweber:1990dv,Leinweber:1992hy} and model expectations.
However as the chiral regime is approached the nonanalytic behavior of
the quenched meson cloud is revealed, enhancing the proton and
suppressing the $\Delta^{+}$ in accord with the expectations of
quenched \chiPT\ \cite{Leinweber:2003ux,Young:2003gd}.  The quenched
artifacts of the $\Delta$ provide an unmistakable signal for the onset
of quenched chiral nonanalytic behavior.

%%%%%%%%%%%%%%%%%%%%%%%%%%%%%%%%%%%%%%%%%%%%%%%%%%%%%%%%%%%%%%%%%%%%%%%%%%%
% \section{Conclusions and Future Work}
\section{CONCLUSIONS}
\label{conclusion}

We have presented the first lattice QCD simulation results for the
electromagnetic form factors of the nucleon, $\Sigma$, $\Xi$ and
$\Delta$ baryons at quark masses light enough to reveal unmistakable
quenched chiral nonanalytic behavior. The non-linear behaviour
observed is in agreement with the predictions of quenched chiral
perturbation theory \cite{Leinweber:2002qb,Young:2003gd}.

This work paves the way for a study of the individual quark sector
contributions to baryon magnetic moments and, in particular, the
strangeness content of the nucleon \cite{DerekLHP03}.

%%%%%%%%%%%%%%%%%%%%%%%%%%%%%%%%%%%%%%%%%%%%%%%%%%%%%%%%%%%%%%%%%%%%%%%%%%%
%\acknowledgements
\vspace*{0.3cm} This work was supported by the Australian Research
Council.  We thank the Australian Partnership for Advanced
Computing (APAC) for generous grants of supercomputer time which have
enabled this project.

%%%%%%%%%%%%%%%%%%%%%%%%%%%%%%%%%%%%%%%%%%%%%%%%%%%%%%%%%%%%%%%%%%%%%%%%%%%

\vfill
\end{document}